\documentclass[epsfig,12pt]{article}
\usepackage{makeidx}
\usepackage{amsmath}
\usepackage{amsfonts}
\usepackage{amssymb}
\usepackage{graphicx}

\input epsf.sty
\newtheorem{theorem}{Theorem}
\newtheorem{acknowledgement}[theorem]{Acknowledgement}

\makeatletter \@addtoreset{equation}{section}
\renewcommand{\theequation}{\thesection.\arabic{equation}}
\input{tcilatex}
\begin{document}

\title{\rightline{\mbox{\small
{LPHE-MS-10-04/CPM-10-04}}} \bigskip \textbf{Graphene, Lattice QFT and
Symmetries}}
\author{L.B Drissi$^{a}$, E.H Saidi$^{a,b,c}$, M. Bousmina$^{a}\medskip $ \\
{\small a. INANOTECH, Institute of Nanomaterials and Nanotechnology, Rabat,
Morocco,}\\
{\small b. LPHE- Modelisation et Simulation, Facult\'{e} des Sciences Rabat,
Morocco}\\
c. {\small Centre of Physics and Mathematics, CPM, Rabat, Morocco}}
\maketitle

\begin{abstract}
{Borrowing ideas from tight binding model, we propose a board class of
Lattice QFT models that are classified by the ADE Lie algebras. In the case
of }${A}_{{N-1}}\simeq su\left( N\right) ${\ series, we show that the
couplings between the quantum states living at the first nearest neighbor
sites of the lattice }$\mathcal{L}_{su\left( N\right) }${\ are governed by
the complex\ fundamental representations }\underline{${{\mathbf{N}}}$}{\ and 
}$\overline{{\mathbf{N}}}${\ of }$su\left( N\right) ${; and the second
nearest neighbor interactions are described by its adjoint \underline{${{%
\mathbf{N}}}$}}$\otimes \overline{{\mathbf{N}}}${. The lattice models
associated with the leading }$su\left( 2\right) ${, }$su\left( 3\right) ${\
and }$su\left( 4\right) ${\ cases are explicitly studied and their fermionic
field realizations are given. It is also shown that the su}$\left( {2}%
\right) ${\ and su}$\left( {3}\right) ${\ models describe respectively the
electronic properties of the acetylene chain and the graphene. It is
established as well that the} energy dispersion of the first nearest
neighbor couplings is completely determined by the ${A}_{{N}}$ roots $%
\mathbf{\alpha }$ through the typical dependence $N/2+\sum_{roots}\cos
\left( \mathbf{k}.\alpha \right) $ with $\mathbf{k}$ the wave vector.{\
Other features such as DE extension and other applications are also
discussed.}\newline
\end{abstract}


\section{Introduction}

Tight Binding Model (TBM) \textrm{\cite{1,2,3,4}} is a particular lattice
QFT \textrm{\cite{5,6,7,8}} modeling couplings between quantum states living
at closed neighboring sites. The interactions are modeled in terms of hops
of crystal particles or holes; and brings together issues from High Energy
Physics and Condensed Matter \textrm{\cite{9A,9B,10,11,12,13,14}}. Despite
the restriction to the first nearest neighbor interactions, the studies
based on TBM have been shown to capture the main information on the physical
properties of lattice systems; like in graphene \textrm{\cite{15,16,17}}
whose basic physical properties have been related to QED in $\left(
2+1\right) $ dimensions; for reviews see \textrm{\cite{9A,9B}, refs therein
and \cite{18,19}}.\newline
In this paper, we use TBM to engineer a board class of lattice QFTs that are
based on ADE Lie algebras \textrm{\cite{191,192}} and their basic
representations \textrm{\cite{193,194}}. These engineered lattice systems
classify the electronic properties of acetylene chain as a $su\left(
2\right) $ model, graphene as a $su\left( 3\right) $ lattice model and may
have application in other fields; in particular in QFT on non commutative
geometry \textrm{\cite{20,21A,21B},} where space time is viewed as a
crystal, and in the special subset of conformal field models based on affine
Kac-Moody invariance\textrm{\ }and vertex operators\textrm{\ \cite{22,23,24}}%
. \newline
To fix the ideas; let us describe briefly the main lines of the construction
in the case of the series $A_{N-1}\simeq su\left( N\right) $ \textrm{\cite%
{231} }which is generated by $\left( N-1\right) $ commuting Cartan
generators $h^{i}$ and $N\left( N-1\right) $ step operators $E^{\pm \beta }$
where the vectors $\mathbf{\beta }=\left( \beta ^{1},...,\beta ^{N-1}\right) 
$ stand for the positive roots of $su\left( N\right) $. As a QFT on crystal,
our $su\left( N\right) $ lattice model involves the two basic following :%
\newline
(\textbf{1}) \emph{the lattice} $\mathcal{L}_{su\left( N\right) }$:\newline
It is made by the superposition of two sublattices $\mathcal{A}_{su\left(
N\right) }$ and $\mathcal{B}_{su\left( N\right) }$ generated by the $%
su\left( N\right) $ simple roots as in eqs(\ref{z1}). This $\left(
N-1\right) $- dimensional lattice extends the \emph{1D} chain and the well
known \emph{2D} honeycomb of graphene corresponding to $N=2$ and $N=3$
respectively; see figures (\ref{ch}), (\ref{co}) for illustration. \newline
Each $\mathbf{r}_{m}$ site of $\mathcal{L}_{su\left( N\right) }$; say $%
\mathbf{r}_{m}\in \mathcal{A}_{su\left( N\right) }$, has $N$ first nearest
neighbors at $\left( \mathbf{r}_{m}+\mathbf{v}_{i}\right) \in \mathcal{B}%
_{su\left( N\right) }$ and $N\left( N-1\right) $ second nearest neighbors at 
$\left( \mathbf{r}_{m}+\mathbf{V}_{ij}\right) \in \mathcal{A}_{su\left(
N\right) }$ with the two remarkable relations%
\begin{equation}
\begin{tabular}{lllll}
{\small 1st nearest} & : & $\mathbf{v}_{0}+\mathbf{v}_{1}+\ldots +\mathbf{v}%
_{N-1}=0$ & , & {\small (a)} \\ 
{\small 2nd nearest} & : & $\mathbf{V}_{ij}=\mathbf{v}_{i}-\mathbf{v}_{j}$ & 
, & {\small (b)}%
\end{tabular}
\label{bti}
\end{equation}%
that respectively have interpretation in terms of the vector weights of the
fundamental and the adjoint representations of $su\left( N\right) $. \newline
Recall that the weight vectors $\mathbf{\mu }_{i}$ (resp. $\mathbf{\beta }%
_{ij}=\mathbf{\mu }_{i}-\mathbf{\mu }_{j}$) of the complex N dimensional
fundamental (resp. adjoint) representations of su$\left( N\right) $ obey the
following relations 
\begin{equation}
\begin{tabular}{lllll}
{\small fundamental} & : & $\mathbf{\mu }_{0}+\mathbf{\mu }_{1}+\ldots +%
\mathbf{\mu }_{N-1}=0$ & , & {\small (a)} \\ 
{\small adjoint} & : & $\mathbf{\beta }_{ij}=\mathbf{\mu }_{i}-\mathbf{\mu }%
_{j}$ & , & {\small (b)}%
\end{tabular}
\label{bt}
\end{equation}%
which are analogous to (\ref{bti}) and so solve them by taking $\mathbf{v}%
_{i}=a\mathbf{\mu }_{i}$ and $\mathbf{V}_{ij}=a\beta _{ij}$ with constant $a$
to be interpreted later on. \newline
Notice that the constraint eqs $\mathbf{\mu }_{0}+\mathbf{\mu }_{1}+\ldots +%
\mathbf{\mu }_{N-1}=0$ and similarly \ $\sum \mathbf{\beta }_{ij}=0$ are
also important from the physical side since they are interpreted in terms of
the conservation of total momentum of the outgoing and incoming waves at
each site of $\mathcal{L}_{su\left( N\right) }$ 
\begin{equation}
\begin{tabular}{ll}
$\mathbf{p}_{1}+\ldots +\mathbf{p}_{N-1}+\mathbf{p}_{0}=0$ & .%
\end{tabular}%
\end{equation}%
(\textbf{2}) \emph{the hamiltonian} $\mathcal{H}_{su\left( N\right) }$%
\newline
Denoting by $F_{\mu _{i}}$ (resp. $G_{\beta _{ij}}$) the field operators
generating the hops of the particles/holes between the site $\mathbf{r}_{m}$
and $\mathbf{r}_{m}+\mathbf{v}_{i}$ (resp. $\mathbf{r}_{m}+\mathbf{V}_{ij}$%
), the proposed hamiltonian $\mathcal{H}_{su\left( N\right) }$ describing
the quantum state couplings up to second nearest neighbors on $\mathcal{L}%
_{su\left( N\right) }$ reads as follows,%
\begin{equation}
\begin{tabular}{lll}
$\mathcal{H}_{su\left( N\right) }=$ & $-t_{1}\left( \dsum\limits_{\substack{ 
\text{weights of }  \\ \text{fund of }su\left( N\right) }}F_{\mu
_{k}}+hc\right) -t_{2}\left( \dsum\limits_{su\left( N\right) \text{ }%
roots}G_{\beta _{ij}}+hc\right) $ & 
\end{tabular}
\label{A1}
\end{equation}%
where the $t_{1}$, $t_{2}$ are hop energies. The fermionic field
realisations of $F_{\mu _{k}}$, $G_{\beta _{ij}}$ are given by eqs(\ref{B1}).%
\newline
The proposed $\mathcal{H}_{_{su\left( N\right) }}$ depends on the $su\left(
N\right) $ algebra representation quantities namely the weight vectors $%
\mathbf{\mu }_{i}$ of the fundamental of $su\left( N\right) $ and its roots $%
\mathbf{\beta }_{ij}$. This property leads a priori to an energy spectrum of 
$\mathcal{H}_{su\left( N\right) }$ completely characterized by the wave
vector $\mathbf{k}$, the weights $\mathbf{\mu }_{i}$ and the roots $\mathbf{%
\beta }_{ij}$; but as we will see the $\mathbf{\mu }_{i}$ dependence is
implicit and appears only through the roots. Such results are also shown to
extend naturally to the $so\left( 2N\right) $ lattice models.\newline
The presentation is as follows: In section 2, we develop our proposal for
the case of lattice models based on $su\left( N\right) $ Lie algebras. In
section 3, we consider the $su\left( 2\right) $ and $su\left( 3\right) $
models describing respectively the electronic properties of the acetylene
chain and graphene. In section 4, we deepen the $su\left( 4\right) $ lattice
model and in section 5 we give conclusion and further comments regarding DE
extension.

\section{The proposal: $su\left( N\right) $ model}

In this section, we develop our proposal by first building the real lattice $%
\mathcal{L}_{su\left( N\right) }$ that is associated with the hamiltonian (%
\ref{A1}) refered to as the su$\left( N\right) $ lattice model. Then, we
give a QFT realization of the field operators $F_{\mu _{i}}$ and $G_{\beta
_{ij}}$ using free fermionic fields on $\mathcal{L}_{su\left( N\right) }$.
We also give the energy dispersion $\varepsilon _{su\left( N\right) }\left( 
\mathbf{k}\right) $ relation in terms of the wave vector $\mathbf{k}$, the
weights $\mathbf{\mu }_{i}$ and the roots $\beta _{ij}$.

\subsection{Building the lattice $\mathcal{L}_{su\left( N\right) }$}

\begin{figure}[tbph]
\begin{center}
\hspace{0cm} \includegraphics[width=12cm]{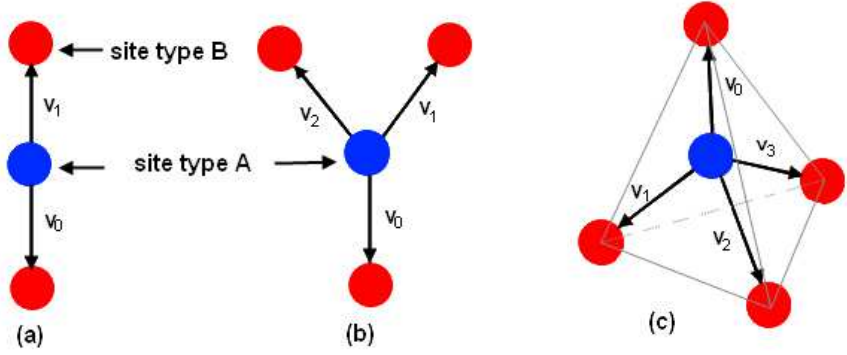}
\end{center}
\par
\vspace{-0.5 cm}
\caption{(\textbf{a}) 1A+2B lattice sites of $\mathcal{L}_{su\left( 2\right)
}$; A-type in blue and B-type in red; the 2B form a $su\left( 2\right) $
doublet. (\textbf{b}) 1A+3B sites of $\mathcal{L}_{su\left( 3\right) }$; the
3B form a $su\left( 3\right) $ triplet. (\textbf{c}) 1A+4B sites of $%
\mathcal{L}_{su\left( 4\right) }$ with 4B sites forming a regular
tetrahedron. }
\label{123}
\end{figure}

The lattice $\mathcal{L}_{su\left( N\right) }$ is a real $\left( N-1\right) $%
- dimensional crystal with two superposed integral sublattices $\mathcal{A}%
_{su\left( N\right) }$ and $\mathcal{B}_{_{su\left( N\right) }}$; each site $%
\mathbf{r}_{\mathbf{m}}$ of these sublattices is generated by the $su\left(
N\right) $ simple roots $\mathbf{\alpha }_{1},...,\mathbf{\alpha }_{N-1}$; 
\begin{equation}
\begin{tabular}{llll}
$\mathbf{r}_{\mathbf{m}}$ & $=$ & $\dsum\limits_{m_{1},...,m_{N-1}}m_{1}%
\mathbf{\alpha }_{1}+m_{2}\mathbf{\alpha }_{2}+...m_{N-1}\mathbf{\alpha }%
_{N-1}$ & ,%
\end{tabular}
\label{z1}
\end{equation}%
with $m_{i}$ integers; for illustration see the schema (a), (b), (c) of the
figure (\ref{123}) corresponding respectively to $N=2,3,4$ and which may be
put in one to one with the $sp^{1}$, $sp^{2}$ and $sp^{3}$ hybridization of
the carbon atom orbitals $2s$ and $2p$.

On each lattice site $\mathbf{r}_{m}$ of $\mathcal{L}_{su\left( N\right) }$;
say of A-type, lives a quantum state $A\left( \mathbf{r}_{m}\right) $
coupled to the nearest neighbor states; in particular the first nearest
states $B\left( \mathbf{r}_{m}+\mathbf{v}_{i}\right) $ and the second
nearest ones $A\left( \mathbf{r}_{m}+\mathbf{V}_{ij}\right) $. \newline
Generally, generic sites in $\mathcal{L}_{su\left( N\right) }$ have the
following properties:\newline
(\textbf{1}) $N$ first nearest neighbors with relative position vectors $%
\mathbf{v}_{i}$ constrained as 
\begin{equation}
\begin{tabular}{lll}
$+\mathbf{v}_{0}+\mathbf{v}_{1}+\ldots +$ $\mathbf{v}_{N-1}$ & $=0$ & ,%
\end{tabular}
\label{21}
\end{equation}%
or equivalently%
\begin{equation}
\begin{tabular}{lll}
$-\mathbf{v}_{0}-\mathbf{v}_{1}-\ldots -$ $\mathbf{v}_{N-1}$ & $=0$ & ,%
\end{tabular}%
\end{equation}%
respectively related with the fundamental \underline{$\mathbf{N}$} and
anti-fundamental $\mathbf{\bar{N}}$ representations of $su\left( N\right) $.
Indeed, by using (\ref{bt}-a), these constraint relations are solved in
terms of the $su\left( N\right) $ weight vectors $\mathbf{\mu }_{i}$ (resp. $%
-\mathbf{\mu }_{i}$) of the fundamental (anti-fundamental) representation as
follows%
\begin{equation}
\begin{tabular}{lllll}
$\mathbf{v}_{i}$ & $=a\mathbf{\mu }_{i}$ & $\equiv $ & $d\frac{\mathbf{\mu }%
_{i}}{\left\Vert \mathbf{\mu }_{i}\right\Vert }$ & ,%
\end{tabular}
\label{22}
\end{equation}%
where $d$ is the relative distance between the closest $\mathcal{L}%
_{su\left( N\right) }$ sites. From the QFT view, this means that the quantum
states at $\mathbf{r}_{m}+\mathbf{v}_{i}$ sites are labeled by the $\mathbf{%
\mu }_{i}$ weights as 
\begin{equation}
\begin{tabular}{lll}
$B\left( \mathbf{r}_{m}+\mathbf{v}_{i}\right) $ & $\equiv B_{\mathbf{\mu }%
_{i}}\left( \mathbf{r}_{m}\right) $ & ,%
\end{tabular}%
\end{equation}%
and so the multiplet%
\begin{equation}
\begin{tabular}{ll}
$\left( 
\begin{array}{c}
|B_{\mathbf{\mu }_{0}}> \\ 
\vdots \\ 
|B_{\mathbf{\mu }_{N-1}}>%
\end{array}%
\right) $ & 
\end{tabular}%
\end{equation}%
transform in the fundamental representation of $su\left( N\right) $ and its
conjugate in the anti-fundamental. \newline
(\textbf{2}) $N\left( N-1\right) $ second nearest neighbors of A-type with
relative position vectors $\mathbf{V}_{ij}$ given by eq(\ref{bti}-b) and
obeying the constraint relation,%
\begin{equation}
\begin{tabular}{lll}
$\dsum\limits_{i,j}\mathbf{V}_{ij}$ & $=0$ & .%
\end{tabular}
\label{cd}
\end{equation}%
This condition is naturally solved by (\ref{bt}-a) and (\ref{22}) showing
that the relative vectors between second nearest neighbors are proportional
to su$\left( N\right) $ roots $\mathbf{\beta }_{ij}$ like 
\begin{equation}
\begin{tabular}{lll}
$\mathbf{V}_{ij}=a\mathbf{\beta }_{ij}$ & , & $\mathbf{\beta }_{ij}=\mathbf{%
\mu }_{i}-\mathbf{\mu }_{j}$%
\end{tabular}
\label{26}
\end{equation}%
and so the condition (\ref{cd}) turns to a $su\left( N\right) $ property on
its adjoint representation labeled by the roots.

\subsection{More on $\mathcal{L}_{su\left( N\right) }$}

To get more insight into the structure of the lattice $\mathcal{L}_{su\left(
N\right) }$, it is useful to recall some basic results on $su\left( N\right) 
$ \textrm{\cite{231}}. \newline
This algebra has $\frac{N\left( N-1\right) }{2}$ positive roots $\mathbf{%
\beta }_{ij}$ with $i>j$, which we denote collectively as $+\beta $, and $%
\frac{N\left( N-1\right) }{2}$ negative ones $-\mathbf{\beta }$ so that the
sum on the total roots is zero%
\begin{equation}
\begin{tabular}{lll}
$\dsum\limits_{\func{positive}\text{ roots}}\mathbf{\beta }+\dsum\limits_{%
\func{negative}\text{ roots}}\mathbf{\beta }$ & $=\mathbf{0}$ & .%
\end{tabular}%
\end{equation}%
This property which captures (\ref{cd}) is precisely the analog of eq(\ref%
{bti}-a) for the case of the the adjoint representation of $su\left(
N\right) $. \newline
Moreover, the $\pm \mathbf{\beta }$ roots have same length $\mathbf{\beta }%
^{2}=2$ and are given by positive/negative integral combinations of the $%
\left( N-1\right) $ simple roots $\mathbf{\alpha }_{1},...,\mathbf{\alpha }%
_{n-1}$ 
\begin{equation}
\begin{tabular}{llll}
$\pm \beta =\pm \dsum\limits_{i}l_{i}\alpha _{i}$ & , & $l_{i}\in Z_{+}$ & .%
\end{tabular}
\label{28}
\end{equation}%
Notice that the simple roots $\mathbf{\alpha }_{i}$ are basic objects in Lie
algebras; they capture several information. In particular, they allow to
define the fundamental weight vectors $\mathbf{\lambda }_{i}$ obeying 
\begin{equation}
\begin{tabular}{ll}
$\mathbf{\lambda }_{i}.\mathbf{\alpha }_{j}=\delta _{ij}$ & ,%
\end{tabular}
\label{29}
\end{equation}%
and give as well the intersection matrix 
\begin{equation}
\begin{tabular}{llllll}
$\mathbf{K}_{ij}=\frac{2\left( \mathbf{\alpha }_{i},\mathbf{\alpha }%
_{j}\right) }{\left( \mathbf{\alpha }_{i},\mathbf{\alpha }_{i}\right) }$ & $%
= $ & $\mathbf{\alpha }_{i}.\mathbf{\alpha }_{j}$ & , & $\mathbf{\alpha }%
_{i}.\mathbf{\alpha }_{i}=2$ & 
\end{tabular}%
\end{equation}%
encoding all data on the Lie algebra properties of $su\left( N\right) $.
This matrix is real symmetric reading as,%
\begin{equation}
\begin{tabular}{ll}
$\mathbf{K}_{ij}=\left( 
\begin{array}{cccccc}
2 & -1 & 0 & \cdots & 0 & 0 \\ 
-1 & 2 & -1 &  & 0 & 0 \\ 
0 & -1 & 2 &  & 0 & 0 \\ 
\vdots &  &  & \ddots &  & \vdots \\ 
0 & 0 & 0 &  & 2 & -1 \\ 
0 & 0 & 0 & \cdots & -1 & 2%
\end{array}%
\right) _{r\times r}$ & ,%
\end{tabular}%
\end{equation}%
with rank $r=\left( N-1\right) $. \newline
Notice also that $su\left( N\right) $ has $\left( N^{2}-1\right) $
dimensions generated by $r$ commuting Cartan operators $h^{1},...,h^{r}$
giving the charge vectors of the $su\left( N\right) $ states; and by the
step ones $E^{\pm \beta }$ allowing to hop between the states of the
representation. These operators obey the commutation relations, 
\begin{equation}
\begin{tabular}{lll}
$\left[ h^{i},h^{j}\right] $ & $=$ & $0$ \\ 
$\left[ h^{i},E^{\beta }\right] $ & $=$ & $\beta ^{i}E^{\beta }$ \\ 
$\left[ E^{-\beta },E^{\beta }\right] $ & $=$ & $\frac{2}{\beta ^{2}}\beta
.h $ \\ 
$\left[ E^{\alpha },E^{\beta }\right] $ & $=$ & $\varepsilon _{\alpha \beta
}E^{\alpha +\beta }\text{ \ \ \ \ if \ }\alpha +\beta \ \text{is a root}$%
\end{tabular}%
\end{equation}%
and are used to construct highest weight state representation (HWR) with
highest state $|\phi _{\lambda }>$ and highest weight vector (dominant
weight) $\mathbf{\lambda }$ solving the following constraint relations%
\begin{equation}
\begin{tabular}{lll}
$E^{+\beta }|\phi _{\lambda }>$ & $=0$ & , \\ 
$h^{i}|\phi _{\mathbf{\lambda }}>$ & $=\lambda ^{i}|\phi _{\mathbf{\lambda }%
}>$ & .%
\end{tabular}%
\end{equation}%
The other $\left( N-1\right) $ states $|\phi _{\mu _{i}}>$ of the
representation are obtained by successive actions on $|\phi _{\lambda }>$ by
the typical monomials $E^{-\beta _{m}}\ldots E^{-\beta _{2}}E^{-\beta _{1}}$%
. \newline
One of these HWRs is precisely the N dimensional fundamental representation
we are interested in here; it has $N$ states,%
\begin{equation}
\begin{tabular}{ll}
$\left( 
\begin{array}{c}
F_{\mu _{0}} \\ 
F_{\mu _{1}} \\ 
\vdots \\ 
F_{\mu _{N-1}}%
\end{array}%
\right) $ & ,%
\end{tabular}%
\end{equation}%
with weight vectors 
\begin{equation}
\begin{tabular}{ll}
$\mathbf{\mu }_{i}=\mathbf{\lambda }-\sum_{l=1}^{i}\mathbf{\beta }_{l}$ & ,%
\end{tabular}
\label{ml}
\end{equation}%
\ satisfying (\ref{bti}-a) with $\mathbf{\mu }_{0}=\mathbf{\lambda }$; and
from which we learn that 
\begin{equation}
\begin{tabular}{ll}
$\mathbf{\mu }_{i+1}-\mathbf{\mu }_{i}$ & ,%
\end{tabular}%
\end{equation}%
is indeed an $su\left( N\right) $ root. For an illustration of (\ref{ml});
see the explicit analysis regarding the $su\left( 4\right) $ lattice model;
in particular eq(\ref{mm}).

\subsection{Fermionic realization of $\mathcal{H}_{su\left( N\right) }$}

Denoting by $A_{\mathbf{r}_{m}}^{\pm }$ (resp. $B_{\mathbf{r}_{m}+\mathbf{v}%
_{i}}^{\pm }$) the local fermionic creation and annihilation operators
satisfying the usual anticommutation relations, the hamiltonian on $\mathcal{%
L}_{su\left( N\right) }$ reads as in (\ref{A1}) with $F_{\mu _{i}}$ and $%
G_{\beta }$ operators given by 
\begin{equation}
\begin{tabular}{lll}
$F_{\mu _{i}}$ & $=\dsum\limits_{r_{m}\in \mathcal{A}_{su\left( N\right)
}}A_{\mathbf{r}_{m}}^{-}B_{\mathbf{r}_{m}+a\mu _{i}}^{+}$ & , \\ 
$G_{\beta }$ & $=\dsum\limits_{r_{m}\in \mathcal{A}_{su\left( N\right)
}}\left( A_{\mathbf{r}_{m}}^{-}A_{\mathbf{r}_{m}+b\beta }^{+}+B_{\mathbf{r}%
_{m}}^{-}B_{\mathbf{r}_{m}+b\beta }^{+}\right) $ & ,%
\end{tabular}
\label{B1}
\end{equation}%
where $\mu _{i}$ are the weight vectors of the fundamental representation of 
$su\left( N\right) $ and $\beta $ a generic root. Notice that the operators $%
F_{\mu _{i}}$ and its adjoint $F_{\mu _{i}}^{\dagger }$ transform
respectively in the fundamental representation and its complex conjugate%
\begin{equation}
\begin{tabular}{llllll}
$F_{\mu _{i}}\sim $ & \underline{$N$} & , & $F_{\mu _{i}}^{\dagger }\sim $ & 
$\overline{N}$ & .%
\end{tabular}%
\end{equation}%
By using Fourier transform of the field operators $A_{\mathbf{r}_{m}}^{\pm }$
and $B_{\mathbf{r}_{m}+\mathbf{v}_{i}}^{\pm }$ namely,%
\begin{equation}
\begin{tabular}{llll}
$A^{\pm }\left( \mathbf{r}_{m}\right) $ & $\sim $ & $\dsum\limits_{\text{%
wave vectors }\mathbf{k}}e^{\pm i\mathbf{k.r}_{m}}\tilde{A}\left( \mathbf{k}%
\right) $ &  \\ 
$B^{\pm }\left( \mathbf{r}_{m}+\mathbf{v}_{i}\right) $ & $\sim $ & $%
\dsum\limits_{\text{wave vectors }\mathbf{k}}e^{\pm i\mathbf{k.}\left( 
\mathbf{r}_{m}+\mathbf{v}_{i}\right) }\tilde{B}\left( \mathbf{k}\right) $ & 
\end{tabular}%
\end{equation}%
we can put the hamiltonian $\mathcal{H}_{su\left( N\right) }$ like the sum
over the $\left( N-1\right) $- dimensional wave vectors $\mathbf{k}$ as
follows, 
\begin{equation}
\begin{tabular}{ll}
$\dsum\limits_{\text{wave vectors }\mathbf{k}}\tilde{H}_{\mathbf{k}%
}^{su\left( N\right) }$ & ,%
\end{tabular}%
\end{equation}%
where $\tilde{H}_{\mathbf{k}}^{su\left( N\right) }$ has dispersion relations
depending, in addition to $\mathbf{k}$, on the weights $\mathbf{\mu }_{i}$,
the roots $\mathbf{\beta }$ and the hop energies t$_{1}$, t$_{2}$. In the
particular case where $t_{2}$ is set to zero; the hamiltonian (\ref{A1})
reduces to the leading term 
\begin{equation}
\begin{tabular}{ll}
$\mathcal{H}_{su\left( N\right) }^{1}=-t_{1}\left( \dsum\limits_{weights%
\text{ }\mu _{i}}F_{\mu _{i}}+hc\right) $ & ,%
\end{tabular}%
\end{equation}%
and its dual Fourier transform simplifies as follows,%
\begin{equation}
\begin{tabular}{lll}
$\tilde{H}_{\mathbf{k}}^{su\left( N\right) }=$ & $\left( \tilde{A}_{\mathbf{k%
}}^{-},B_{\mathbf{k}}^{-}\right) \left( 
\begin{array}{cc}
0 & \varepsilon _{su\left( N\right) }\left( \mathbf{k}\right) \\ 
\overline{\varepsilon _{su\left( N\right) }\left( \mathbf{k}\right) } & 0%
\end{array}%
\right) \left( 
\begin{array}{c}
A^{+} \\ 
B^{+}%
\end{array}%
\right) $ & 
\end{tabular}%
\end{equation}%
with 
\begin{equation}
\begin{tabular}{ll}
$\varepsilon _{su\left( N\right) }\left( \mathbf{k}\right) =\dsum\limits_{%
\text{weight vectors }\mu _{i}}e^{ia\mathbf{k}.\mathbf{\mu }_{i}}$ & .%
\end{tabular}%
\end{equation}%
From these relations, we can compute the dispersion energies of the
"valence" and "conducting" bands by diagonalizing $\tilde{H}_{\mathbf{k}%
}^{su\left( N\right) }$. These energies are given by $\pm \left\vert
\varepsilon _{su\left( N\right) }\left( \mathbf{k}\right) \right\vert $ with,%
\begin{equation}
\begin{tabular}{ll}
$\left\vert \varepsilon _{su\left( N\right) }\left( \mathbf{k}\right)
\right\vert =t_{1}\sqrt{N+2\dsum\limits_{i<j=0}^{N-1}\cos \left[ a\mathbf{k}%
.\left( \mathbf{\mu }_{i}\mathbf{-\mu }_{j}\right) \right] }$ & .%
\end{tabular}
\label{di}
\end{equation}%
Notice that $\left\vert \varepsilon _{su\left( N\right) }\left( \mathbf{k}%
\right) \right\vert $ depends remarkably in the difference of the weights $%
\mathbf{\mu }_{i}\mathbf{-\mu }_{j}$; which according to (\ref{28}) is just
the sum over $su\left( N\right) $ roots $\sum_{l=i}^{j}\mathbf{\beta }_{l}$
with $\mathbf{\beta }_{l}=\mathbf{\mu }_{l}\mathbf{-\mu }_{l+1}$. It follows
then that dispersion energies for the first nearest neighbors depend on the
wave vector $\mathbf{k}$ and the roots of $su\left( N\right) $. 
\begin{equation}
\begin{tabular}{ll}
$\left\vert \varepsilon _{su\left( N\right) }\left( \mathbf{k}\right)
\right\vert =t_{1}\sqrt{N+2\dsum\limits_{i<j=0}^{N-1}\cos \left(
a\sum_{l=i}^{j}\mathbf{k}.\mathbf{\beta }_{l}\right) }$ & .%
\end{tabular}
\label{de}
\end{equation}%
This result is expected from group theory view since $\left\vert \varepsilon
_{su\left( N\right) }\right\vert ^{2}=\varepsilon _{su\left( N\right) }%
\overline{\varepsilon _{su\left( N\right) }}$ should be put in
correspondence with the tensor product of the fundamental representation 
\underline{$N$} and its complex conjugate $\bar{N}$ 
\begin{equation}
\begin{tabular}{llllll}
\underline{$N$}$\otimes \bar{N}$ & $=$ & $I_{id}\oplus adj_{SU\left(
N\right) }$ & , & $tr\left( I_{id}\right) =N$ & ,%
\end{tabular}%
\end{equation}%
giving the adjoint representation of $U\left( N\right) \simeq U\left(
1\right) \times SU\left( N\right) $.\newline
Notice finally that eq(\ref{de}) may be further explicited by first
expressing $\mathbf{\beta }_{l}$ in terms of simple roots as in (\ref{28});
that is $\mathbf{\beta }_{l}=\sum_{m=1}^{N-1}M_{ml}\alpha _{m}$ with $M_{nl}$
integers. Then expand the wave vector as $\mathbf{k=}\sum_{n=1}^{N-1}q_{n}%
\mathbf{\lambda }_{n}$ with $q_{n}$ real number; and use (\ref{29}) to put (%
\ref{de}) in the following handleable form%
\begin{equation}
\begin{tabular}{ll}
$\left\vert \varepsilon _{su\left( N\right) }\left( \mathbf{k}\right)
\right\vert =t_{1}\sqrt{N+2\dsum\limits_{i<j=0}^{N-1}\cos \left(
a\dsum\limits_{l=i}^{j}\dsum\limits_{n=1}^{N-1}q_{n}M_{nl}\right) }$ & .%
\end{tabular}%
\end{equation}%
Below, we consider explicit examples.

\section{Leading $su\left( N\right) $ lattice models}

In this section, we illustrate the above study on the leading examples $%
N=2,3 $. These two lattice models describe the electronic properties of the
delocalized electrons in the infinite acetylene type chain and graphene.

\subsection{the $su\left( 2\right) $ model}

\begin{figure}[tbph]
\begin{center}
\hspace{0cm} \includegraphics[width=10cm]{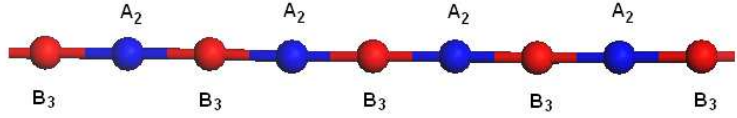}
\end{center}
\par
\vspace{-0.5 cm}
\caption{the {\protect\small lattice }$\mathcal{L}_{su\left( 2\right) }$%
{\protect\small \ given by the superposition of two sublattices }$\mathcal{A}%
_{su\left( 2\right) }${\protect\small \ (in blue) and }$\mathcal{B}%
_{su\left( 2\right) }$ {\protect\small (in red). The atoms may be thought of
as carbons in the }${\protect\small sp}^{{\protect\small 1}}${\protect\small %
\ hybridization state.}}
\label{ch}
\end{figure}

In this case, the lattice $\mathcal{L}_{su\left( 2\right) }$ depicted in the
figure (\ref{ch}) is a one dimensional chain with coordinate positions $%
x_{m}=ma$ where $a$ is the site spacing and $m$ an integer.

Each site of $\mathcal{L}_{su\left( 2\right) }$ has two first nearest
neighbors forming an $su\left( 2\right) $ doublet; and two second nearest
ones associated with the two roots $\pm \alpha $ of $su\left( 2\right) $ in
agreement with the generic result summarized in the table,%
\begin{equation}
\begin{tabular}{l|l|l|l|l}
{\small nearest neighbors} & $su\left( N\right) $ & $su\left( 2\right) $ & $%
su\left( 3\right) $ & $su\left( 4\right) $ \\ \hline
{\small \ \ \ \ \ \ \ \ \ \ \ first } & $N$ & $2$ & $3$ & $4$ \\ 
{\small \ \ \ \ \ \ \ \ \ \ \ second} & $N\left( N-1\right) $ & $2$ & $6$ & $%
12$ \\ \hline
\end{tabular}%
\end{equation}%
In this $N=2$ model, eqs(\ref{bti}) read as%
\begin{equation}
\begin{tabular}{lll}
$\mathbf{v}_{0}+\mathbf{v}_{1}=0$ & , & (a) \\ 
$\mathbf{V}_{01}=\mathbf{v}_{0}-\mathbf{v}_{1}$ & , & (b)%
\end{tabular}%
\end{equation}%
and are solved by the fundamental weights $\mu _{0}=+\frac{1}{2},$ $\mu
_{1}=-\frac{1}{2}$ of the su$\left( 2\right) $ fundamental representation;
i.e the isodoublet. \newline
The hamiltonian $\mathcal{H}_{su\left( 2\right) }$ of this model is given by 
\begin{equation}
\begin{tabular}{lll}
$\mathcal{H}_{su\left( 2\right) }$ & $=-t_{1}\left( F_{+\frac{1}{2}}+F_{-%
\frac{1}{2}}+hc\right) -t_{2}\left( G_{+1}+G_{-1}+hc\right) $ & .%
\end{tabular}%
\end{equation}%
where $t_{1}$ and $t_{2}$ are hop energies. The fermionic field realization
of the $F_{\pm \frac{1}{2}}$ and $G_{\pm 1}$ operators read in terms of the
creation and annihilation $A_{\mathbf{x}_{m}}^{\pm },$ $B_{\mathbf{x}%
_{m}}^{\pm }$ as follows%
\begin{equation}
\begin{tabular}{lll}
$F_{\pm \frac{1}{2}}$ & $=\dsum\limits_{m}A_{\mathbf{x}_{m}}^{-}B_{\mathbf{x}%
_{m}\pm a}^{+}$ & , \\ 
$G_{\pm 1}$ & $=\dsum\limits_{m}\left( A_{\mathbf{x}_{m}}^{-}A_{\mathbf{x}%
_{m}\pm 2a}^{+}+B_{\mathbf{x}_{m}}^{-}B_{\mathbf{x}_{m}\pm 2a}^{+}\right) $
& .%
\end{tabular}%
\end{equation}%
Moreover, substituting $N=2$ in (\ref{di}), we get the dispersion energy%
\begin{equation}
\left\vert \varepsilon _{su\left( 2\right) }\left( k\right) \right\vert
=t_{1}\sqrt{2+2\cos \left( 2ak\right) }
\end{equation}%
which is also equal to $2t_{1}\cos \left( ka\right) $ and from which we read
that the $\left\vert \varepsilon _{su\left( 2\right) }\left( k\right)
\right\vert $ zeros take place for the wave vectors $k_{n}=\pm \frac{\pi }{2a%
}$ $\func{mod}$ $\frac{2\pi }{a}$.

\subsection{the $su\left( 3\right) $ model and graphene}

\begin{figure}[tbph]
\begin{center}
\hspace{0cm} \includegraphics[width=6cm]{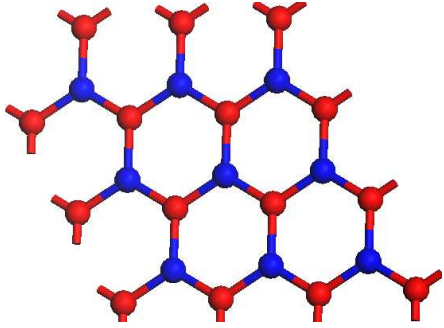}
\end{center}
\par
\vspace{-1cm}
\caption{{\protect\small sublattices A (in blue) and B (in red) of the
honeycomb. The atoms may be thought of as carbons in the }$sp^{%
{\protect\small 2}}${\protect\small \ hybridization state.}}
\label{co}
\end{figure}

The lattice $\mathcal{L}_{su\left( 3\right) }$ is precisely the $2D$
honeycomb of graphene; it is given by the superposition of two sublattices $%
\mathcal{A}_{su\left( 3\right) }$ and $\mathcal{B}_{su\left( 3\right) }$ as
in the figure (\ref{co}).

Each site $\mathbf{r}_{m}$ in $\mathcal{L}_{su\left( 3\right) }$ has \emph{3}
first nearest neighbors with relative vectors $\mathbf{v}_{i}$; and \emph{6}
second neighbors $\pm \mathbf{V}_{ij}=\pm \varepsilon _{ijk}V_{k}$ in one to
one correspondence with the vector weights of the $su\left( 3\right) $
fundamental representation and its roots. We have 
\begin{equation}
\begin{tabular}{llll}
$\mathbf{v}_{i}=a\mathbf{\mu }_{i}$ & , & $\pm \mathbf{V}_{i}=\pm a\mathbf{%
\alpha }_{i}$ & ,%
\end{tabular}
\label{DR}
\end{equation}%
where $a=d\sqrt{\frac{3}{2}}$. To fix the ideas, we give below the explicit
expressions of the weight vectors $\mathbf{\mu }_{1},$ $\mathbf{\mu }_{2},$ $%
\mathbf{\mu }_{0}$ and the roots $\mathbf{\alpha }_{1}$, $\mathbf{\alpha }%
_{2}$, $\mathbf{\alpha }_{3}$, 
\begin{equation}
\begin{tabular}{llllll}
$\mathbf{\mu }_{1}=(\frac{\sqrt{2}}{2},\frac{\sqrt{6}}{6})$ & , & $\mathbf{%
\mu }_{2}=(-\frac{\sqrt{2}}{2},\frac{\sqrt{6}}{6})$ & , & $\mathbf{\mu }%
_{0}=-(0,\frac{\sqrt{6}}{3})$ & , \\ 
$\mathbf{\alpha }_{1}=(\sqrt{2},0)$ & , & $\mathbf{\alpha }_{2}=(-\frac{%
\sqrt{2}}{2},\frac{\sqrt{6}}{2})$ & , & $\mathbf{\alpha }_{3}=(\frac{\sqrt{2}%
}{2},\frac{\sqrt{6}}{2})$ & ,%
\end{tabular}
\label{C1}
\end{equation}%
from which we learn that $\mathbf{\alpha }_{3}=\mathbf{\alpha }_{1}+\mathbf{%
\alpha }_{2}$ as it should be. In addition to the feature $\mathbf{\mu }.%
\mathbf{\alpha \in }\mathbb{Z}$, these vectors obey the constraint relations 
\begin{equation}
\begin{tabular}{lll}
$\dsum\limits_{i=0}^{2}\mathbf{v}_{i}=0$ & , & $\dsum\limits_{\func{positive}%
\text{ roots}}\mathbf{\alpha }+\dsum\limits_{\func{negative}\text{ roots}}%
\mathbf{\alpha }=0$%
\end{tabular}%
\end{equation}%
together with the following remarkable relations 
\begin{equation}
\begin{tabular}{llllll}
$\mathbf{\mu }_{0}-\mathbf{\mu }_{1}=\mathbf{\alpha }_{3}$ & , & $\mathbf{%
\mu }_{1}-\mathbf{\mu }_{2}=\mathbf{\alpha }_{1}$ & , & $\mathbf{\mu }_{2}%
\mathbf{-\mu }_{0}=\mathbf{\alpha }_{2}$ & ,%
\end{tabular}%
\end{equation}%
Substituting $N=3$ in (\ref{di}) and using the above equations, we get the
dispersion energy%
\begin{equation}
\begin{tabular}{ll}
$\left\vert \varepsilon _{su\left( 3\right) }\left( \mathbf{k}\right)
\right\vert =t_{1}\sqrt{3+2\left[ \cos a\mathbf{k.\alpha }_{1}+\cos a\mathbf{%
k.\alpha }_{2}+\cos a\mathbf{k.\alpha }_{3}\right] }$ & ,%
\end{tabular}%
\end{equation}%
depending on the wave vector $\mathbf{k}=\left( k_{x},k_{y}\right) $ and the 
$su\left( 3\right) $ roots.

\section{the $su\left( 4\right) $ lattice model}

\begin{figure}[tbph]
\begin{center}
\hspace{0cm} \includegraphics[width=6cm]{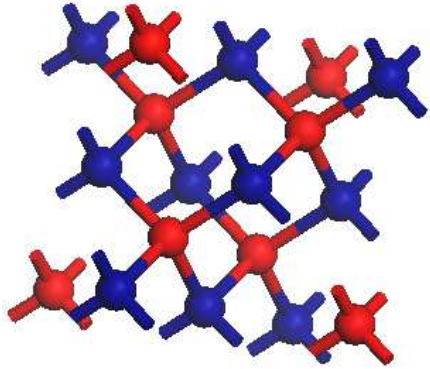}
\end{center}
\par
\vspace{-1cm}
\caption{the lattice $\mathcal{L}_{su\left( 4\right) }$ with {\protect\small %
sublattices }$\mathcal{A}_{su\left( 4\right) }${\protect\small \ (in blue)
and }$\mathcal{B}_{su\left( 4\right) }${\protect\small \ (in red). Each atom
has \emph{4} first nearest neighbors, forming a tetrahedron, and \emph{12}
second nearest ones.}}
\label{4}
\end{figure}

To illustrate the general properties of the $su\left( 4\right) $ model; we
first give some basic features on the connection between $\mathcal{L}%
_{su\left( 4\right) }$, depicted by the figure (\ref{4}), and the $su\left(
4\right) $ representations. Then, we develop a dynamical model based on the
crystal $\mathcal{L}_{su\left( 4\right) }$. 

\subsection{structure property of $\mathcal{L}_{su\left( 4\right) }$}

The lattice $\mathcal{L}_{su\left( 4\right) }$ is a 3-dimensional crystal;
it is made by the superposition of two isomorphic, but shifted, sublattices $%
\mathcal{A}_{su\left( 4\right) }$ and $\mathcal{B}_{su\left( 4\right) }$
following the same logic as in the case of the honeycomb which may be
recovered by projection on a 2D plane. \newline
Each site $\mathbf{r}_{m}$ in $\mathcal{L}_{su\left( 4\right) }$ has \emph{4}
first nearest neighbors at $\left( \mathbf{r}_{m}+\mathbf{v}_{i}\right) $
forming the vertices of a regular tetrahedron. A way to parameterize the
relative positions $\mathbf{v}_{i}$ with respect to the central position at $%
\mathbf{r}_{m}$ is to embed the tetrahedron inside a cube; in this case we
have:%
\begin{equation}
\begin{tabular}{llll}
$\mathbf{v}_{1}=\frac{d}{\sqrt{3}}\left( -1,-1,+1\right) $ & , & $\mathbf{v}%
_{2}=\frac{d}{\sqrt{3}}\left( -1,+1,-1\right) $ &  \\ 
$\mathbf{v}_{3}=\frac{d}{\sqrt{3}}\left( +1,-1,-1\right) $ & , & $\mathbf{v}%
_{0}=\frac{d}{\sqrt{3}}\left( +1,+1,+1\right) $ & 
\end{tabular}
\label{vi}
\end{equation}%
Clearly these vectors satisfy the constraint relation (\ref{bti}-a).
Moreover, having these expressions, we can also build the explicit positions
of the \emph{12} second nearest neighbors; these are given by eq(\ref{bti}%
-b); but are completely generated by the following basis vectors 
\begin{equation}
\begin{tabular}{lllll}
$\mathbf{R}_{1}=\frac{d}{\sqrt{3}}\left( 2,2,0\right) $ & , & $\mathbf{R}%
_{2}=\frac{d}{\sqrt{3}}\left( 0,-2,2\right) $ & , & $\mathbf{R}_{3}=\frac{d}{%
\sqrt{3}}\left( -2,2,0\right) $%
\end{tabular}
\label{RR}
\end{equation}%
that are related to $\mathbf{V}_{ij}$ as follows,%
\begin{equation}
\begin{tabular}{ll}
$\mathbf{R}_{i}=\mathbf{V}_{\left( i-1\right) i}$ & 
\end{tabular}%
\end{equation}%
For later use, it is interesting to notice the two following:\newline
(\textbf{a}) the $\mathbf{R}_{i}$ vector basis have the following
intersection matrix 
\begin{equation}
\mathbf{R}_{i}.\mathbf{R}_{j}=\frac{4d^{2}}{3}\mathbf{K}_{ij}  \label{kij}
\end{equation}%
with $\mathbf{K}_{ij}$ and its inverse as 
\begin{equation}
\begin{tabular}{llll}
$\mathbf{K}_{ij}=\left( 
\begin{array}{ccc}
2 & -1 & 0 \\ 
-1 & 2 & -1 \\ 
0 & -1 & 2%
\end{array}%
\right) $ & , & $\mathbf{K}_{ij}^{-1}=\frac{1}{4}\left( 
\begin{array}{ccc}
3 & 2 & 1 \\ 
2 & 4 & 2 \\ 
1 & 2 & 3%
\end{array}%
\right) $ & 
\end{tabular}%
\end{equation}%
(\textbf{b}) using (\ref{vi}) and (\ref{RR}), it is not difficult to check
the following special relation linking the $\mathbf{R}_{i}$'s and $\mathbf{v}%
_{0}$,%
\begin{equation}
\begin{tabular}{ll}
$\frac{3}{4}\mathbf{R}_{1}+\frac{2}{4}\mathbf{R}_{2}+\frac{1}{4}\mathbf{R}%
_{3}=\mathbf{v}_{0}$ & .%
\end{tabular}
\label{v0}
\end{equation}%
In fact this relation is a $su\left( 4\right) $ Lie algebra property
following from $\mathbf{K}_{ij}^{-1}$ and (\ref{29}). \newline
Concerning the vector positions of the remaining \emph{9} second neighbors, 
\emph{3} of them are given by $-\mathbf{R}_{1},-\mathbf{R}_{2},-\mathbf{R}%
_{3}$ and the other \emph{6} by the linear combinations%
\begin{equation}
\begin{tabular}{llllll}
$+\mathbf{R}_{4}=\mathbf{V}_{02}$ & $=+R_{1}+R_{2}$ & , & $+\mathbf{R}_{5}=%
\mathbf{V}_{13}$ & $=+R_{2}+R_{3}$ &  \\ 
$-\mathbf{R}_{4}=\mathbf{V}_{20}$ & $=-R_{1}-R_{2}$ & , & $-\mathbf{R}_{5}=%
\mathbf{V}_{31}$ & $=-R_{2}-R_{3}$ &  \\ 
$+\mathbf{R}_{6}=\mathbf{V}_{03}$ & $=+R_{1}+R_{2}+R_{3}$ & , & $-\mathbf{R}%
_{6}=\mathbf{V}_{30}$ & $=-R_{1}-R_{2}-R_{3}$ & 
\end{tabular}
\label{R}
\end{equation}%
From this construction, it follows that generic positions $\mathbf{r}_{%
\mathbf{m}}^{A}\equiv \mathbf{r}_{\mathbf{m}}$ and $\mathbf{r}_{\mathbf{m}%
}^{B}$ in the $\mathcal{A}_{su\left( 4\right) }$ and $\mathcal{B}_{su\left(
4\right) }$ sublattices are given by%
\begin{equation}
\begin{tabular}{llllll}
$\mathcal{A}_{su_{4}}$ & : & $\mathbf{r}_{\mathbf{m}}$ & $=$ & $m_{1}\mathbf{%
R}_{1}+m_{2}\mathbf{R}_{2}+m_{3}\mathbf{R}_{3}$ & , \\ 
$\mathcal{B}_{su_{4}}$ & : & $\mathbf{r}_{\mathbf{m}}^{B}$ & $=$ & $\mathbf{r%
}_{\mathbf{m}}+\mathbf{v}$ & ,%
\end{tabular}
\label{mr}
\end{equation}%
where $\mathbf{m}=\left( m_{1}\mathbf{,}m_{2}\mathbf{,}m_{3}\right) $ is an
integer vector and where the shift vector $\mathbf{v=r}_{\mathbf{m}}^{B}-%
\mathbf{r}_{\mathbf{m}}^{A}$ is one of $\mathbf{v}_{i}$'s in (\ref{vi}).%
\newline
Regarding the connection between $\mathcal{L}_{su\left( 4\right) }$ and the $%
su\left( 4\right) $ Lie algebra representations, we distinguish two kinds of
relations:\newline
(\textbf{1}) \emph{a relation involving the roots}\newline
From (\ref{kij}), it follows that the basis vectors $\mathbf{R}_{1}$, $%
\mathbf{R}_{2}$, $\mathbf{R}_{3}$ may be interpreted in terms of the simple
roots $\mathbf{\alpha }_{1},$ $\mathbf{\alpha }_{2},$ $\mathbf{\alpha }_{3}$
of the $su\left( 4\right) $ Lie algebra. More precisely, we have%
\begin{equation}
\begin{tabular}{lllll}
$\mathbf{R}_{1}=\frac{2d}{\sqrt{3}}\mathbf{\alpha }_{1}$ & , & $\mathbf{R}%
_{2}=\frac{2d}{\sqrt{3}}\mathbf{\alpha }_{2}$ & , & $\mathbf{R}_{3}=\frac{2d%
}{\sqrt{3}}\mathbf{\alpha }_{3}$%
\end{tabular}
\label{ro}
\end{equation}%
from which we learn that the matrix $K_{ij}=\mathbf{\alpha }_{i}.\mathbf{%
\alpha }_{j}$ is indeed the Cartan matrix of $su\left( 4\right) $. \newline
We also have the following relations showing that the position vectors of
the second nearest neighbors are indeed in one to one correspondence with
the roots of $su\left( 4\right) $,%
\begin{equation}
\begin{tabular}{lllll}
$\mathbf{R}_{4}=\frac{2d}{\sqrt{3}}\left( \mathbf{\alpha }_{1}+\mathbf{%
\alpha }_{2}\right) $ & $,$ & $\mathbf{R}_{5}=\frac{2d}{\sqrt{3}}\left( 
\mathbf{\alpha }_{2}+\mathbf{\alpha }_{3}\right) $ & , & $\mathbf{R}_{6}=%
\frac{2d}{\sqrt{3}}\left( \mathbf{\alpha }_{3}+\mathbf{\alpha }_{1}\right) $%
\end{tabular}
\label{ri}
\end{equation}%
together with their opposites.\newline
(\textbf{1}) \emph{a relation involving the fundamental representation of su}%
$\left( \emph{4}\right) $\newline
As described for the generic $su\left( N\right) $, the four relative vectors 
$\mathbf{v}_{i}$ are, up to a global scale factor, nothing but the four
weight vectors $\mathbf{\mu }_{i}$ of the fundamental representation of $%
su\left( 4\right) $. The highest weight $\mathbf{\lambda }$ of this
representation, which we set as $\mathbf{\mu }_{0}$, is precisely given by
eq(\ref{v0}). Thus, we have%
\begin{equation}
\begin{tabular}{lllllll}
$\mathbf{v}_{0}=\frac{2d}{\sqrt{3}}\mathbf{\mu }_{0}$ & , & $\mathbf{v}_{1}=%
\frac{2d}{\sqrt{3}}\mathbf{\mu }_{1}$ & , & $\mathbf{v}_{2}=\frac{2d}{\sqrt{3%
}}\mathbf{\mu }_{2}$ & , & $\mathbf{v}_{3}=\frac{2d}{\sqrt{3}}\mathbf{\mu }%
_{3}$%
\end{tabular}
\label{am}
\end{equation}%
that obey obviously the constraint relation%
\begin{equation}
\begin{tabular}{ll}
$\mathbf{\mu }_{0}+\mathbf{\mu }_{1}+\mathbf{\mu }_{2}+\mathbf{\mu }_{3}=%
\mathbf{0}$ & .%
\end{tabular}
\label{m}
\end{equation}%
This constraint equation is a typical vector relation for highest weight
representations of Lie algebras; it extends the well known $su\left(
2\right) $ ones whose leading terms are%
\begin{equation}
\begin{tabular}{l|l}
${\small su}\left( {\small 2}\right) $ & {\small sum over weights} \\ \hline
{\small doublet} & $\frac{1}{2}-\frac{1}{2}=0$ \\ 
{\small triplet} & $1+0-1=0$ \\ 
{\small quartet} & $\frac{3}{2}+\frac{1}{2}-\frac{3}{2}-\frac{1}{2}=0$ \\ 
\hline
\end{tabular}%
\end{equation}%
To exhibit more explicitly the constraint relation (\ref{m}), it is
interesting to express the weight vectors $\mathbf{\mu }_{i}$ in terms of
the simple roots of su$\left( 4\right) $. We have 
\begin{equation}
\begin{tabular}{llll}
$\mathbf{\mu }_{0}$ & $\mathbf{=}$ & $+\frac{3}{4}\mathbf{\alpha }_{1}+\frac{%
2}{4}\mathbf{\alpha }_{2}+\frac{1}{4}\mathbf{\alpha }_{3}$ &  \\ 
$\mathbf{\mu }_{1}$ & $\mathbf{=}$ & $-\frac{1}{4}\mathbf{\alpha }_{1}+\frac{%
2}{4}\mathbf{\alpha }_{2}+\frac{1}{4}\mathbf{\alpha }_{3}$ &  \\ 
$\mathbf{\mu }_{2}$ & $\mathbf{=}$ & $-\frac{1}{4}\mathbf{\alpha }_{1}-\frac{%
2}{4}\mathbf{\alpha }_{2}+\frac{1}{4}\mathbf{\alpha }_{3}$ &  \\ 
$\mathbf{\mu }_{3}$ & $\mathbf{=}$ & $-\frac{1}{4}\mathbf{\alpha }_{1}-\frac{%
2}{4}\mathbf{\alpha }_{2}-\frac{3}{4}\mathbf{\alpha }_{3}$ & 
\end{tabular}
\label{mm}
\end{equation}%
This analysis teaches us as well two basic things:\newline
First the number $N_{1}$ of the first nearest neighbors in the lattice $%
\mathcal{L}_{su\left( 4\right) }$ is related to the dimension of the
fundamental representation of su$\left( 4\right) $%
\begin{equation}
N_{1}=\dim \left( \underline{\mathbf{4}}\right) =4
\end{equation}%
This means that the QFT on this lattice should capture some data on $%
su\left( 4\right) $.\newline
Second, the number $N_{2}$ of the second nearest neighbors in $\mathcal{L}%
_{su\left( 4\right) }$ is also related to a su$\left( 4\right) $ quantity
namely,%
\begin{equation}
N_{2}=\dim \left[ {\small su}\left( 4\right) \right] -\text{rank}\left[ 
{\small su}\left( 4\right) \right] =15-3
\end{equation}%
These two Lie algebra numbers may be used as an algorithm to extend this
construction to the case of the $so\left( 2N\right) $ and the $E_{6}$, $%
E_{7} $, $E_{8}$ exceptional simply laced Lie algebras.

\subsection{dynamical vacancy on lattice: a toy model}

We begin by noting that, as far as the electronic properties are concerned,
the schemas (a), (b), (c) of figure (\ref{123}) may be respectively
associated with the $sp^{1}$, $sp^{2}$ and $sp^{3}$ hybridizations of the
atom orbitals; i.e:%
\begin{equation}
\begin{tabular}{l|l|l}
{\small \ \ \ figures} & {\small hybridization} & {\small example of
molecules} \\ \hline
\ \ {\small (\ref{123}-a)} & ${\small \ \ \ \ sp}^{1}$ & {\small acetylene}
\\ 
\ \ {\small (\ref{123}-b)} & ${\small \ \ \ \ sp}^{2}$ & {\small graphene}
\\ 
\ \ {\small (\ref{123}-c)} & ${\small \ \ \ \ sp}^{3}$ & {\small diamond} \\ 
\hline
\end{tabular}%
\end{equation}%
In the two first examples, the atoms have delocalized pi- electrons that
capture the electronic properties of the lattice atoms and have the
following dispersion relation, 
\begin{equation}
\begin{tabular}{ll}
$\left\vert \varepsilon _{su\left( N\right) }\left( \mathbf{k}\right)
\right\vert =t_{1}\sqrt{N+2\dsum\limits_{i<j=0}^{N-1}\cos \left[ a\mathbf{k}%
.\left( \mathbf{\mu }_{i}\mathbf{-\mu }_{j}\right) \right] }$ & 
\end{tabular}%
\end{equation}%
with $N=2,3$. \newline
In the case of $sp^{3}$, the atoms have no delocalized pi-electrons; they
only have strongly correlated sigma- electrons which make the electronic
properties of systems based on $\mathcal{L}_{su\left( 4\right) }$ different
from those based on $\mathcal{L}_{su\left( 3\right) }$ and $\mathcal{L}%
_{su\left( 2\right) }$. \newline
However, as far as tight binding model idea is concerned, one may consider
other applications; one of which concerns the following toy model describing
a system based on the lattice $\mathcal{L}_{su\left( 4\right) }$ with
dynamical vacancy sites.

\emph{Toy model} \newline
This is a lattice QFT on the $\mathcal{L}_{su\left( 4\right) }$ with
dynamical particles and vacancies. The initial state of the system
correspond to the configuration where the sites of the sublattice $\mathcal{A%
}_{su\left( 4\right) }$ are occupied by particles and those of the
sublattice $\mathcal{B}_{su\left( 4\right) }$ are unoccupied.%
\begin{equation}
\begin{tabular}{l|l|l}
{\small sublattice} & {\small initial configuration} & {\small quantum state}
\\ \hline
$\mathcal{A}_{su_{4}}$ & {\small particles at }$\mathbf{r}_{m}$ & $\mathbf{A}%
\left( \mathbf{r}_{m}\right) $ \\ 
$\mathcal{B}_{su_{4}}$ & {\small vacancy at }$\mathbf{r}_{m}+\mathbf{v}$ & $%
\mathbf{B}\left( \mathbf{r}_{m}+\mathbf{v}\right) $ \\ 
&  &  \\ \hline
\end{tabular}%
\end{equation}%
Then, the particles (vacancies) start to move towards the neighboring sites
with movement modeled by hop to first nearest neighbors. \newline
Let $A\left( \mathbf{r}_{m}\right) $ and $B\left( \mathbf{r}_{m}+\mathbf{v}%
_{i}\right) $ be the quantum states describing the particle at $\mathbf{r}%
_{m}$ and the vacancy at $\mathbf{r}_{m}+\mathbf{v}_{i}$ respectively. Let
also $A_{\mathbf{r}_{m}}^{\pm }$ and $B_{\mathbf{r}_{m}+\mu _{i}}^{\pm }$ be
the corresponding creation and annihilation operators. The hamiltonian
describing the hop of the vacancy/particle to the first nearest neighbors is
given by%
\begin{equation}
\begin{tabular}{ll}
$\mathcal{H}_{su_{4}}=$ & $-t_{1}\left( \dsum\limits_{i=0}^{3}A_{\mathbf{r}%
_{m}}^{-}B_{\mathbf{r}_{m}+\mu _{i}}^{+}+hc\right) $%
\end{tabular}%
\end{equation}%
where the $\mu _{i}$'s are the weight vectors of the fundamental of $%
su\left( 4\right) $. By performing Fourier transforms of the $A_{\mathbf{r}%
_{m}}^{\pm }$, $B_{\mathbf{r}_{m}+\mu _{i}}^{\pm }$ field operators , we end
with the dispersion energy%
\begin{equation}
\begin{tabular}{ll}
$\left\vert \varepsilon _{su_{4}}\left( k\right) \right\vert =t_{1}\sqrt{%
4+2\dsum\limits_{i<j}\cos \left( \mathbf{k.V}_{ij}\right) }$ & ,%
\end{tabular}%
\end{equation}%
with $\mathbf{V}_{ij}$ as in (\ref{v0}-\ref{R}). This relation depends on
the wave vector $\mathbf{k}=\left( k_{x},k_{y}\right) $ and the $su\left(
4\right) $ roots.

\section{Conclusion and comments}

\begin{figure}[tbph]
\begin{center}
\hspace{0cm} \includegraphics[width=4cm]{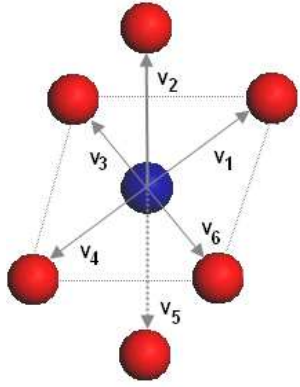}
\end{center}
\par
\vspace{-1cm}
\caption{the lattice $\mathcal{L}_{so\left( 6\right) }${\protect\small ;
each (blue) atom has \emph{6} first nearest neighbors constituting an
octahedron and transforming in the vector representation of so}$\left( 
{\protect\small 6}\right) .$ }
\label{br}
\end{figure}

In the present paper, we have used TBM to engineer a board class of systems
that are based on $su\left( N\right) $ Lie algebras and their basic
representations. These engineered systems classify the acetylene chain and
graphene as $su\left( 2\right) $ and $su\left( 3\right) $ models
respectively. Our construction may have other applications; in particular in
QFT on non commutative geometry where space time is a lattice and in lattice
QFT for condensed matter as exemplified by the dynamical vacancy/particle
toy model introduced in subsection \emph{4.2}.\newline
Our $su\left( N\right) $ lattice models relies on the two following basic
ingredients:\newline
(\textbf{1}) the $N$ first nearest neighbors in $\mathcal{L}_{su\left(
N\right) }$ with positions $\mathbf{v}_{i}=a\mathbf{\mu }_{i}$ and wave
functions $\phi _{\mathbf{\mu }_{i}}\left( \mathbf{r}_{m}\right) $ transform
in the fundamental representation of $su\left( N\right) $. The $\mathbf{v}%
_{i}$'s and the weight vectors $\mathbf{\mu }_{i}$ are as in eqs(\ref{bti}%
-a, \ref{bt}-a ),\newline
(\textbf{2}) the $N\left( N-1\right) $ second nearest neighbors in $\mathcal{%
L}_{su\left( N\right) }$ with positions $\mathbf{V}_{ij}=\mathbf{v}_{i}-%
\mathbf{v}_{j}$ are, up to a global scale factor, given by the $N\left(
N-1\right) $ roots $\mathbf{\beta }_{ij}$ of $su\left( N\right) $.\newline
Using these features, we have constructed a $su\left( N\right) $ lattice
model with hamiltonian $\mathcal{H}_{su\left( N\right) }$ (\ref{A1}); it
involves operator fields $F_{\mathbf{\mu }_{i}}$ and $G_{\beta }$
transforming respectively in the fundamental and adjoint representations of $%
su\left( N\right) $. This symmetry captures basic data on the energy
spectrum of $\mathcal{H}_{su\left( N\right) }$ as shown by the dispersion
energy given by the formulae (\ref{di}-\ref{de}). \newline
Our proposal may be extended to the other DE simply laced Lie algebras. In
the case of $D_{N}\sim so\left( 2N\right) $ for instance, the lattice $%
\mathcal{L}_{so\left( 2N\right) }$ is N- dimensional generated by the simple
roots $\mathbf{\alpha }_{1},\ldots ,\mathbf{\alpha }_{N}$ with matrix
intersection%
\begin{equation}
\begin{tabular}{llll}
$\mathbf{\alpha }_{i}.\mathbf{\alpha }_{j}$ & $=$ & $\left( 
\begin{array}{cccccccc}
2 & -1 & 0 & \cdots & 0 & 0 & 0 & 0 \\ 
-1 & 2 & -1 & \cdots & 0 & 0 & 0 & 0 \\ 
0 & -1 & 2 &  & 0 & 0 & 0 & 0 \\ 
\vdots & \vdots &  & \ddots &  &  & \vdots & \vdots \\ 
0 & 0 & 0 &  & 2 & -1 & 0 & 0 \\ 
0 & 0 & 0 &  & -1 & 2 & -1 & -1 \\ 
0 & 0 & 0 & \cdots & 0 & -1 & 2 & -1 \\ 
0 & 0 & 0 & \cdots & 0 & -1 & -1 & 2%
\end{array}%
\right) $ & 
\end{tabular}%
\end{equation}%
Here also, the $\mathcal{L}_{so\left( 2N\right) }$ lattice is made by the
superposition of two sublattices $\mathcal{A}_{so\left( 2N\right) }$ and $%
\mathcal{B}_{so\left( 2N\right) }$. For each site at $\mathbf{r}_{m}$, we
have $2N$ first nearest neighbors at $\mathbf{r}_{m}+\mathbf{v}_{I}$ with
relative vectors $\mathbf{v}_{I}$, which may be split as, 
\begin{equation}
\begin{tabular}{llllll}
$\mathbf{v}_{i}$ & , & $\mathbf{v}_{i+N}=-\mathbf{v}_{i}$ & , & $1\leq i\leq
N$ & ,%
\end{tabular}%
\end{equation}%
satisfying the following constraint relation 
\begin{equation}
\begin{tabular}{llll}
$\dsum\limits_{I=1}^{2N}\mathbf{v}_{I}$ & $=$ & $\dsum\limits_{i=1}^{N}%
\left( \mathbf{v}_{i}+\mathbf{v}_{i+N}\right) =0$ & 
\end{tabular}%
\end{equation}%
which extend the su$\left( N\right) $ one given by (\ref{bti}-a). In the
example of so$\left( 6\right) $, the first nearest neighbors form an
octahedron as depicted in figure (\ref{br}),

The relative vectors $\mathbf{v}_{I}$ transform in the \underline{$\mathbf{2N%
}$} vector representation of $so\left( 2N\right) $; they allow to define the
second nearest neighbors by help of (\ref{bti}-b) which reads in the $%
so\left( 2N\right) $ case as follows 
\begin{equation}
\begin{tabular}{llll}
$\mathbf{V}_{IJ}$ & $=$ & $\left\{ 
\begin{array}{c}
\pm \left( \mathbf{v}_{i}-\mathbf{v}_{j}\right) \\ 
\pm \left( \mathbf{v}_{i}+\mathbf{v}_{j}\right)%
\end{array}%
\right. $ & 
\end{tabular}
\label{vv}
\end{equation}%
These $\mathbf{V}_{IJ}$'s should be put in one to one correspondence with
the $2N\left( N-1\right) $ roots of $so\left( 2N\right) $. Recall that $%
so\left( 2N\right) $ is $N\left( 2N-1\right) $- dimensional and has rank $N$%
; that is $N$ simple roots $\mathbf{\alpha }_{i}$ which read in terms of the
weight vectors of the \underline{$\mathbf{2N}$} representation like $\mathbf{%
\alpha }_{i}=\mathbf{\mu }_{i}-\mathbf{\mu }_{i+1}$ and $\mathbf{\alpha }%
_{N}=\mathbf{\mu }_{N-1}+\mathbf{\mu }_{N}$. The generic roots are given by $%
\pm \left( \mathbf{\mu }_{i}\pm \mathbf{\mu }_{j}\right) $ with $1\leq
i,j\leq N$ and should be compared with (\ref{vv}). In the end notice that
the dispersion energy for the first nearest couplings reads as%
\begin{equation}
\left\vert \varepsilon _{so\left( 2N\right) }\left( \mathbf{k}\right)
\right\vert =t_{1}\sqrt{2N+2\dsum\limits_{i<j=0}^{N}\cos \left[ a\mathbf{k}%
.\left( \mathbf{\mu }_{i}\mathbf{-\mu }_{j}\right) \right]
+2\dsum\limits_{i<j=0}^{N}\cos \left[ a\mathbf{k}.\left( \mathbf{\mu }_{i}%
\mathbf{+\mu }_{j}\right) \right] }
\end{equation}%
and, like in the $su\left( N\right) $ case (\ref{di}), it also depends on
the so$\left( 2N\right) $ roots $\beta _{ij}=\mathbf{\mu }_{i}\mathbf{\pm
\mu }_{j}$.

\ \ \ \newline
We end this conclusion by making one more comment concerning some related
works on fermions living on a \emph{4d} \emph{hyperdiamond} lattice $%
\mathcal{H}_{4}$ which has been used in lattice QCD \textrm{\cite{BBTW}; see
also \cite{KM}-\cite{KM4} for extensions\footnote{%
we thank the referee for pointing to us the relationship between our
lattices $\mathcal{L}_{su\left( D+1\right) }$ based on Lie algebras,
constructed in section 2, and the D-dimensional hyperdiamond \ $\mathcal{H}%
_{D}$ described in this section.}}. It is interesting to note that the \emph{%
4d} \emph{hyperdiamond} lattice $\mathcal{H}_{4}$ used in \textrm{\cite{BBTW}%
} is precisely $\mathcal{L}_{su\left( 5\right) }$; and the higher
dimensional diamonds $\mathcal{H}_{N}$ given in \textrm{\cite{KM} }are
exactly the $\mathcal{L}_{su\left( N+1\right) }$ lattices we have discussed
in section 2. Moreover, several features obtained for the \emph{N}%
-dimensional hyperdiamond $\mathcal{H}_{N}$ with $N\geq 2$ are just
algebraic relations on the weight and root systems of the $su\left(
N+1\right) $ Lie algebra with discrete symmetries generated by Weyl group
transformations given by the $\mathcal{S}_{N+1}$ permutation group. This is
the case for instance of the remarkable relation, 
\begin{equation}
\begin{tabular}{llll}
$\cos \vartheta _{ij}=\frac{\mathbf{a}_{i}.\mathbf{a}_{j}}{\left\Vert 
\mathbf{a}_{i}\right\Vert .\left\Vert \mathbf{a}_{j}\right\Vert }=\frac{1}{2}
$ & , & $i\neq j=1,...,N$ & ,%
\end{tabular}
\label{cos}
\end{equation}%
derived in \textrm{\cite{KM} where} $\vartheta _{ij}=\left( \widehat{\mathbf{%
a}_{i},\mathbf{a}_{j}}\right) $ are the angles between the primitive vectors 
$\mathbf{a}_{i}=\mathbf{e}_{i}-\mathbf{e}_{5}$ of the lattice $\mathcal{H}%
_{N}$ and where the $\mathbf{a}_{i}$'s stand for the generators of the
sublattices $\mathcal{A}_{N}$ (resp. $\mathcal{B}_{N}$) of the \emph{N}%
-dimensional hyperdiamond $\mathcal{H}_{N}$. Notice that eq(\ref{cos}) is
independent on lattice dimension and on the orientation of the primitive
vectors.\textrm{\ }A way to prove the universality of this relation is to
relate it with basic relations of Lie algebras. A lengthy, but
straightforward, analysis shows that the \emph{5} bond vectors $\mathbf{e}%
_{i}$ and the \emph{4} primitive $\mathbf{a}_{i}$ used in \textrm{\cite{BBTW}
}are respectively related to the \emph{5} weight vectors $\mathbf{\mu }_{i}$
of the fundamental representation of $su\left( 5\right) $ and its \emph{4}
simple roots $\mathbf{\alpha }_{i}$ as follows%
\begin{equation}
\begin{tabular}{llll}
$\mathbf{e}_{i}=\frac{\sqrt{5}}{2}\mathbf{\mu }_{i}$, & $\mathbf{\mu }_{i}.%
\mathbf{\mu }_{i}=\frac{4}{5}$, & $\mathbf{\mu }_{i}.\mathbf{\mu }_{j}=-%
\frac{1}{5},$ & 
\end{tabular}%
\end{equation}%
and%
\begin{equation}
\begin{tabular}{ll}
$\mathbf{a}_{1}$ & $=-\frac{\sqrt{5}}{2}\mathbf{\alpha }_{1}$ \\ 
$\mathbf{a}_{2}$ & $=-\frac{\sqrt{5}}{2}\left( \mathbf{\alpha }_{1}+\mathbf{%
\alpha }_{2}\right) $ \\ 
$\mathbf{a}_{3}$ & $=-\frac{\sqrt{5}}{2}\left( \mathbf{\alpha }_{1}+\mathbf{%
\alpha }_{2}+\mathbf{\alpha }_{3}\right) $ \\ 
$\mathbf{a}_{4}$ & $=-\frac{\sqrt{5}}{2}\left( \mathbf{\alpha }_{1}+\mathbf{%
\alpha }_{2}+\mathbf{\alpha }_{3}+\mathbf{\alpha }_{4}\right) $%
\end{tabular}%
\end{equation}%
From these realizations, the constraint equation $\sum \mathbf{e}_{i}=0$
corresponds to the property $\sum \mathbf{\mu }_{i}=0$; see also (\ref{bt}).
Moreover eq(\ref{cos}) can be read in terms of the $su\left( 5\right) $
Cartan matrix $K_{ij}=\frac{2\mathbf{\alpha }_{i}.\mathbf{\alpha }_{j}}{%
\mathbf{\alpha }_{i}.\mathbf{\alpha }_{i}}$.

\begin{acknowledgement}
: L.B Drissi would like to thank ICTP for the Associationship program.
\end{acknowledgement}

\end{document}